\documentclass[12pt]{iopart}
\usepackage{iopams}
\usepackage{graphicx}
\begin{document}

\title{Beaming Selection and   SN-GRB-Jets Evolution}

\author{D.Fargion, D. D'Armiento\address{Physics Department and INFN, Rome University 1, Sapienza, Ple.A.Moro 2}%
              email: daniele.fargion@roma1.infn.it}

{\centerline{\bf Abstract} \small \vskip1mm \noindent
   After a decade of Fireball reign  there is a hope for thin collimated Jet to solve the Supernova-GRB
   mystery
\maketitle

\section{Introduction: Ten years of SN-GRB- Jets}

\begin{figure}[h]
\begin{center}
\includegraphics[width=4in, height=2.1in]{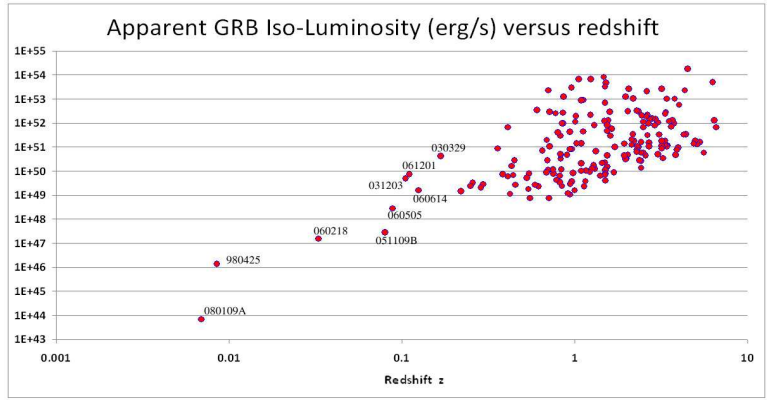}
 \caption{The GRB X-ray luminosity versus redshift updated  to 2008 included. The apparent GRB law Luminosity-red-shift evolution in a quadratic power, is mostly due, (in lower regions) to the quadratic distance cut-off and (in higher regions) to the rarer beaming in axis occurring mostly by largest samples and cosmic volumes. The extreme near SN-GRB discover (2-3 sources) imply a huge population of such events (tens of millions at each year) in cosmic distances (z$>$1), a rate comparable with  SN ones: a large fraction of SN do contain a GRB and the Jet persistence and spreading increases the GRB detection } \label{fig1}
\end{center}
\end{figure}


After a decade (since 25 April 08) we know that  Supernovas may often
   (if not always) contain a Jet. However the Luminosity GRB spread is huge and apparently
   it is evolving just around us. The geometry selection is the cause.   Once on line a blazing jet rises as a GRB.
   The wide apparent GRB power $10^{45} erg s^{-1}\leftrightarrow10^{54} erg s^{-1}$
   may be explained by comparable SN-power beamed jet collimated
   within $\frac{\Delta\Omega}{\Omega}\approx 10^{-7}\leftrightarrow10^{-9}$ thin solid angle.
   (On the contrary (until now) most popular fireball one-shoot explosion
   declared a mild cone beam $\frac{\Delta\Omega}{\Omega}\approx 10^{-2}\leftrightarrow10^{-3}$).
   Because the rarety to be on axis the most distant (within largest sample)
   may contain the most collimated and apparently brightest and hardest sources.
   The inner jet cone are blazing more collimated and their timing variability appear the faster.
   The nearest GRB would appear as the softer and less powerful ones because
   by statistical arguments (near distances-small volumes- small samples) are seen mostly off axis at widest cone periphery. Our
   most updated table of (almost) all GRBs   shows the evidence for such $apparent$
   redshift-Luminosity   evolution. The evolution starts from nearest distances (27 Mpc) up to largest redshift.
      We claim that it is just a selection due to the geometry and due to the detection threshold.
   The persistent  activity of such thin  beamed gamma jet may be powered
    by either a BH (Black Holes) or Pulsars of few Solar Masses. Its activity is longer than the
    observation and duration of GRB, enlarging the GRB solid angle spread and explaining the
    $apparent$ beaming-probability rarety.  Late stages of these jets may be decayed and be suppressed hiding  the isotropic SN echoes
     avoiding easy repetitive GRB signals, even if rare re-brightening took place. Late GRB-SN-Jet may appear
  (if bent and beamed to us)  as a short GRB or a long orphan GRB (depending on jet angular velocity and
  view angle) because its blazing occurred much later than SN observable signal.
   XRF are also such off-axis  viewing of such late  jets.
  Galactic sources as SGR and AXP, at latest stages and minimal output,
   may shine in similar ways as GRB.
   These precessing and spinning $\gamma$ jet are originated by relativistic electron pairs (tens-hundred GeV) by
Inverse Compton and-or Synchrotron Radiation at pulsars or micro-quasars sources.
These Jets are most powerful at Supernova birth. The trembling of the thin jet
bent by random magnetic fields near source, explains
naturally the observed erratic multi-explosive structure of different GRBs.
 The jets are often spinning and precessing (by binary companion or
inner disk asymmetry) and decaying by power law $\approx\frac{t_o}{t}$ on
time scales $t_o$ a few hours; commonly they  keep staying inside the
observer cone view only a few seconds duration times (GRB); the
jet is thinner in gamma and wider in X band by relativistic laws. This explain the
wider and longer X GRB (as well as optical) afterglow duration and the otherwise unexplained presences of
seldom X-ray precursors. This also explain the rarity ($0.5\%$) of GeV Gamma Burst found only at highest red-shift.
The far distance edges (z>4-6)  may dilute and hide the GRB-SN isotropic bump signature. The same selection explains
 the (up to day) absence of GRB hundreds GeV or TeV gamma GRB signals events. Corresponding hundreds GeVs muon neutrinos from GRBs in $km^3$
  mostly correlated with far redshift GRB arrival directions may be too rare to be observed. We suggest the opposite : to follow up the
  hundred GeV  muon neutrino arrival directions with X-ray Telescope in Space to disentangle by X-afterglow eventual hidden gamma burst.
  In conclusion  in our view GRB-SN jet are not the most explosive events in the Universe, but among the most collimated ones.
\section{A list of  GRB puzzles for Fireball}
Why GRBs are so spread in their total  energy, (above 8-9 orders of
magnitude) and in their peak energy  following the so-called Amati
correlation\cite{Amati}? Does the Amati law imply more and more
new GRB source families? Why, as shown below the GRB energy is not a
constant but a growing function (almost quadratic) of the
red-shift?

\begin{figure}[h]
\begin{center}
\includegraphics[width=3in]{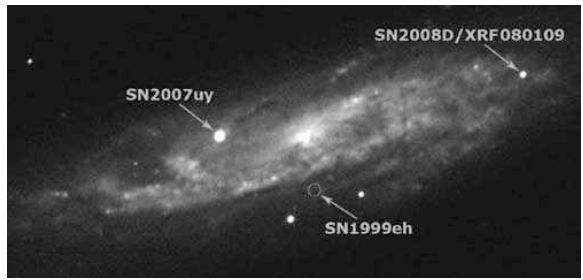}
 \caption{The rare NGC 2770 twice SN within a week time: the XRF080109-SN2008D has deep meaning even for most sceptic theorist} \label{fig1}
 \end{center}
\end{figure}

\begin{figure}[h]
\begin{center}
\includegraphics[width=2.8in]{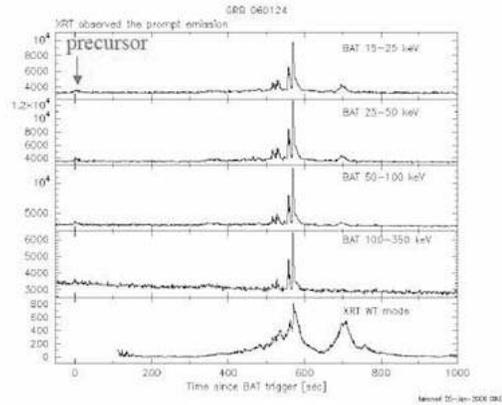}
 \caption{Left: The last GRB070616  long X-ray life: the puzzling ten-minute X-Ray precursor in GRB060124 followed by a huge (apparent) explosive burst (in Fireball model). In our precession model we are in and out an already active  persistent Jet axis} \label{fig1}
\end{center}
\end{figure}


Why are the harder and more variable GRBs (\cite{Lazzati, Fa99})
found at higher redshifts contrary to expected Hubble law? Why
does the output power of GRB vary in a range (\cite{Fa99}) of 8-9
orders of magnitudes with the most powerful events residing at the
cosmic edges (\cite{Yo2004}), see Fig.\ref{fig1}? Why has it been possible to find in
the local universe (at distances 40-150 Mpc just a part over a
million of cosmic space) at least two nearby events (GRB980425 at
$z=0.008$ and recent GRB060218 at $z=0.03$) while most GRBs should
be located at largest volumes, at $z\geq 1$ (\cite{Fa99})? Why are
these two nearby GRBs so much under-luminous (\cite{Fa99})? Why
are their evolution times so slow and smooth respect cosmic ones?
Why do their afterglows show so many bumps and re-brightening as
the well-known third nearest event, GRB030329, if they are
one-shot explosive event? Indeed why do not many GRB curves show
monotonic decay (an obvious consequence of a one-shot explosive
event), rather they often show sudden re-brightening or bumpy
afterglows at different time scales and wavelengths (\cite{Stanek,
DaF03}) - see e.g. GRB050502B\cite{Falcone}? Why have there been a
few GRBs and SGRs whose spectra and time structure are almost
identical if their origin is so different (beamed explosion for
GRB versus isotropic magnetar)\cite{Fa99, Woo99}? How can a jetted
fireball (with an opening angle of $5^o$-$10^o$ and solid angle as
wide as $0.1 sr.$) release an energy-power $10^{50}$ erg$ s^{-1}$
nearly 6 orders of magnitude more energetic than $10^{44}$ erg$
s^{-1}$ the corresponding isotropic SN? Why there is not a more
democratic energy redistribution (or energy equipartition).
\begin{figure}[t]
\includegraphics[width=1.6in]{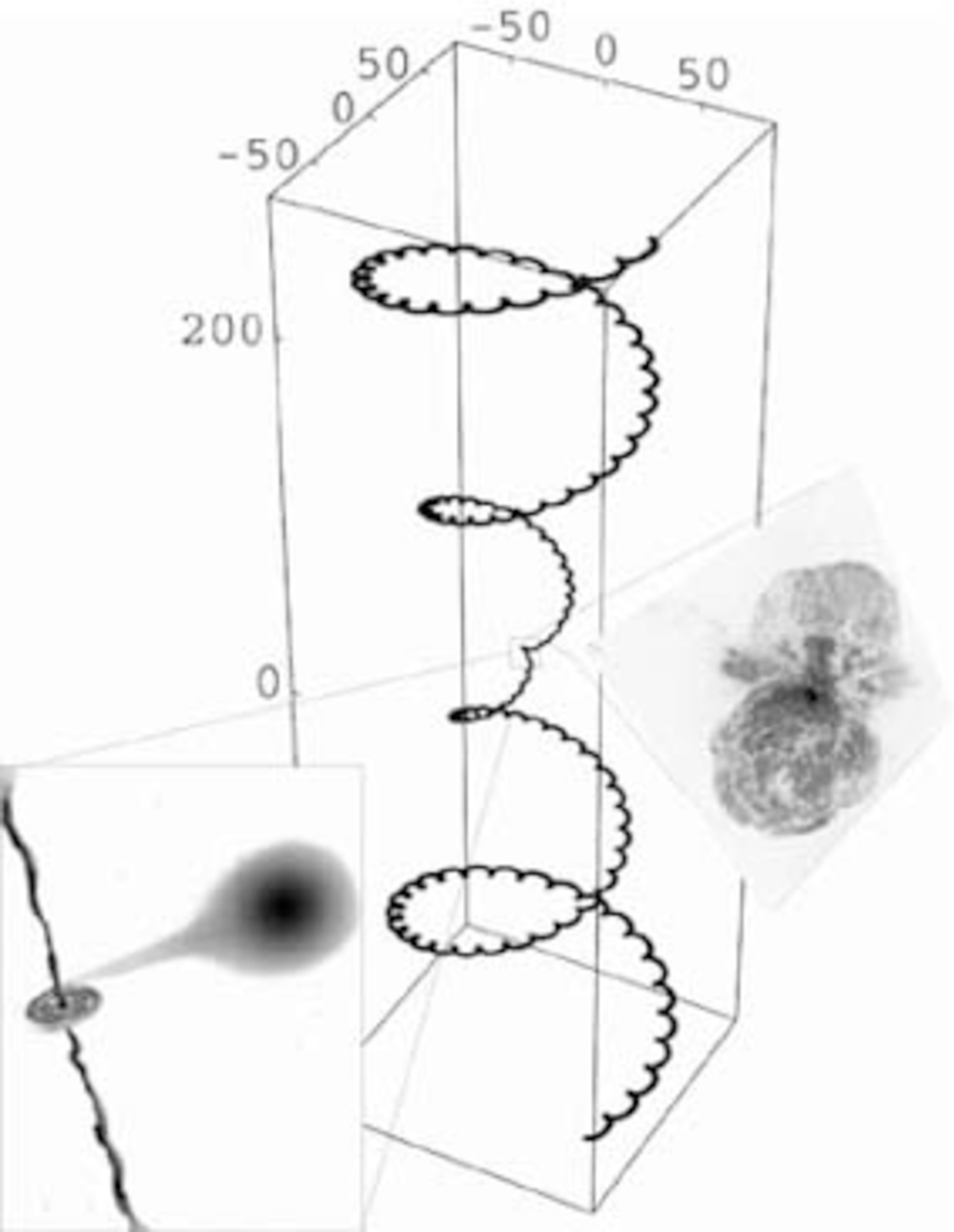}
\includegraphics[width=0.9in]{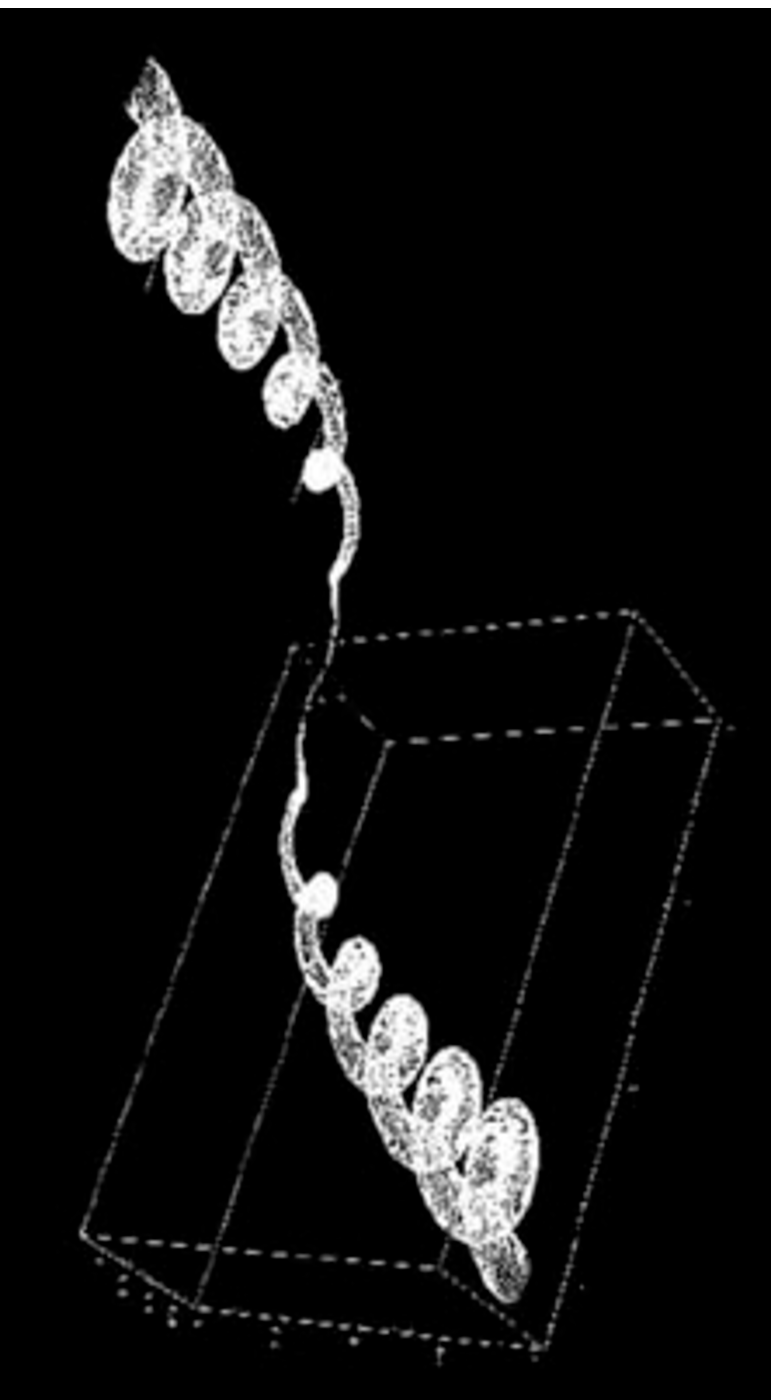}
\includegraphics[width=0.5in]{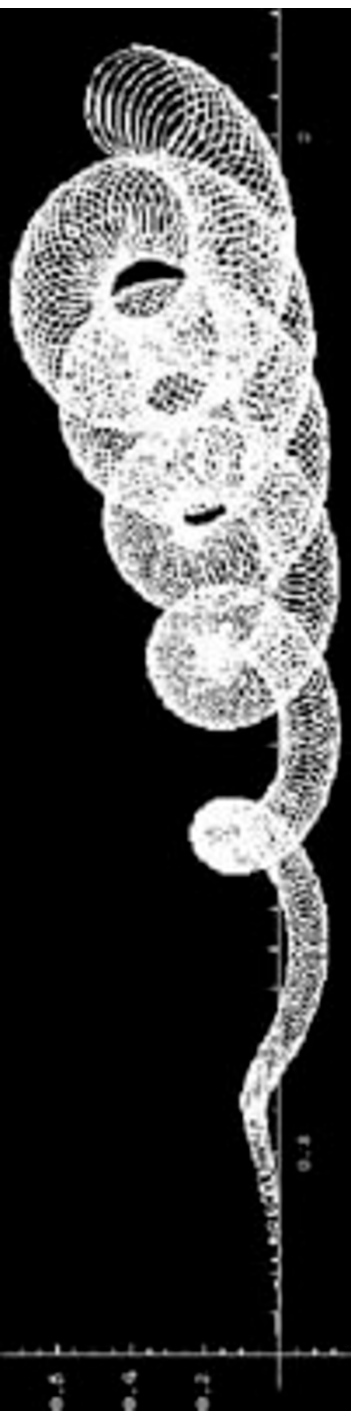}
\includegraphics[width=1.6in]{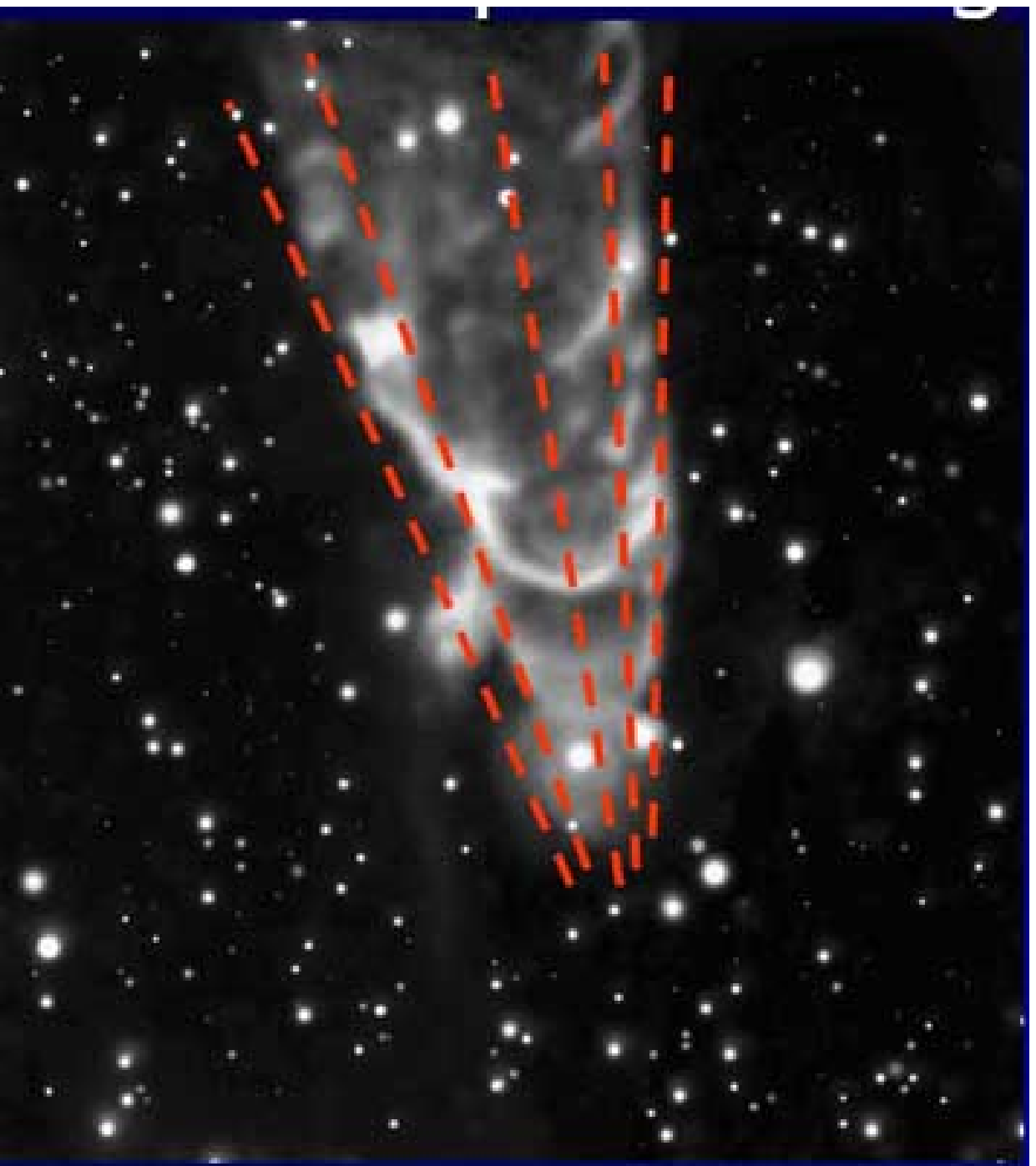}
\includegraphics[width=1.3in]{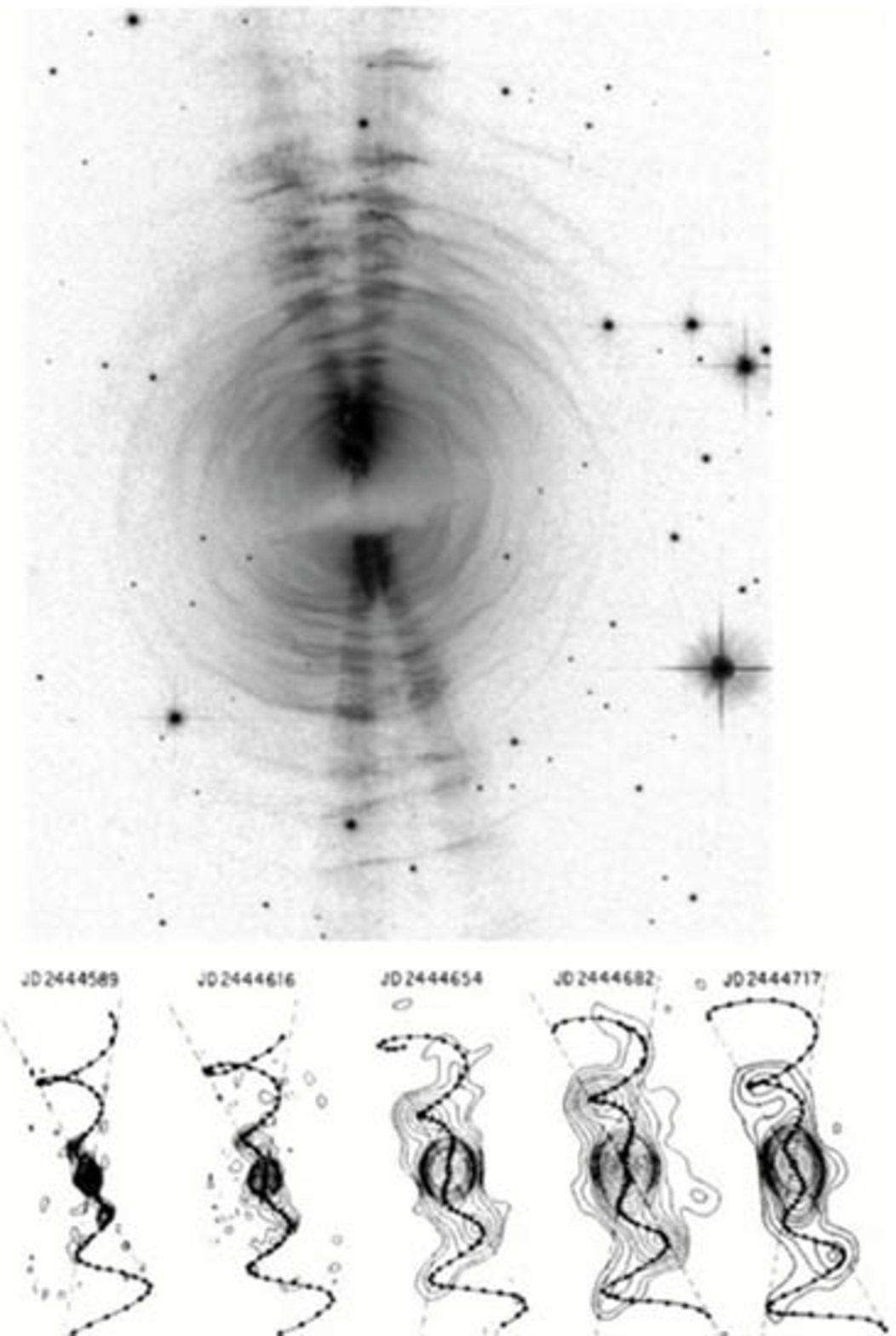}
\caption{From the left to the right: A possible \emph{3D}
structure view of the precessing jet obtained with a precessing and spinning, gamma jet; at its center the "explosive" SN-like source for a GRB ( or a steady binary system, like Eta-Carina, for a SGRs) where an accretion disc around a compact object, powers a thin collimated precessing jet. In the two center figures, the
\emph{3D} and the projected \emph{2D} of such  similar precessing
Jet. In the right fourth panel we show an Herbig Haro-like object
HH49, whose spiral jets are describing, in our opinion, at a lower energy scale,
such precessing Jets . Finally the well known micro-quasars SS-433, its radio evolution spirals in lowest right figure;
 over it stands the egg-nebula shape suggesting a similar inner conical
section of a twin precessing Jet } \label{fig2}
\end{figure}
 How Fireball Jet Model  may fine tune
multi-shells around a GRB in order to produce tuned shock
explosions and re-brightening with no opacity within minutes,
hours, days time-distances from the source(\cite{DaF03})?  How can
some ($\sim6\%$) of the GRBs (or a few SGRs) survive the "tiny"
(but still extremely powerful) pre-explosion of its
\textit{precursor} without any consequences for the source, and
then explode catastrophically few minutes later? In such a
scenario, how could the very recent GRB060124 (at redshift
$z=2.3$) be preceded by a 10 minutes precursor, and then being
able to produce multiple bursts hundreds of times brighter?  Why is the short GRB050724 able to bump and
re-bright a day after the main burst\cite{Campana}?  In this connection
why are the GRB021004 light curves (from X to radio) calling for
an early and late energy injection? Why rarest and historical
GRB940217, highest energetic event, could held more than $5000$s long.?
\begin{figure}[t]
\begin{center}
\includegraphics[width=2.4in]{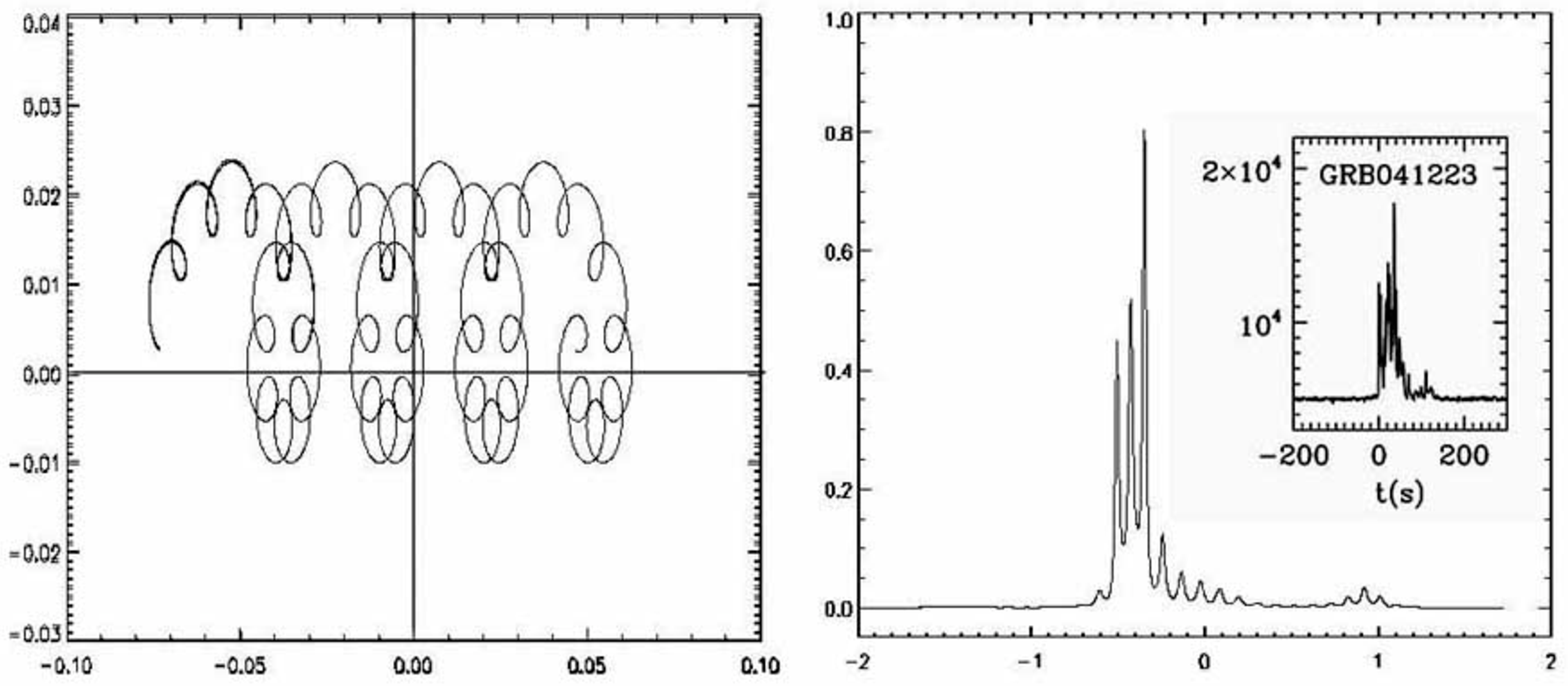}
\caption{The possible simple beam track of a precessing jet to
observer located at origin. On the left, observer stays in (0.00 ;
0.00); the progenitor electron pair jet (leading by
IC\cite{FaSa98} to a gamma jet) has here a Lorentz factor of a
thousand and consequent solid angle at $\sim\mu$ sr. Its
consequent blazing light curve corresponding to such a similar
outcome observed in GRB041223.}\label{fig3}
\end{center}
\end{figure}

Once these major questions are addressed and (in our opinion)
mostly solved by our precessing gamma jet model, a final  question
still remains, calling for a radical assumption on the thin
precessing gamma jet: how can an ultra-relativistic electron beam
(in any kind of Jet models) survive the SN background and dense
matter layers and escape in the outer space while remaining
collimated? Such questions are ignored in most Fireball models
that try to fit the very different GRB afterglow light curves with
shock waves on tuned shells and polynomial ad-hoc curves around
the GRB event. Their solution forces us more and more toward a
unified  precessing Gamma Jet model feeded by the PeV-TeV lepton
showering (about UHE showering beam see analogous ones\cite{Fa97,
Fa00-04}) into $\gamma$ discussed below. As we will show, the thin
gamma precessing jet is indeed made by a chain of primary
processes (PeV muon pair bundles decaying into electrons and then
radiating via synchrotron radiation), requiring an inner
ultra-relativistic jet inside the source.

\section{Blazing Spinning and
Precessing jets in GRBs} The huge GRBs luminosity (up to $10^{54}$
erg s$^{-1}$) may be due to a high collimated on-axis blazing jet,
powered by a Supernova output; the gamma jet is made by
relativistic synchrotron radiation and the inner the jet the
harder and the denser is its output. The harder the photon energy,
the thinner is the jet opening angle. The hardest and shortest
core Gamma event occur at maximal apparent luminosity once the jet
is beamed in inner axis. The jets whole lifetime, while decaying
in output, could survive as long as thousands of years, linking
huge GRB-SN jet apparent Luminosity to more modest SGR relic Jets
(at corresponding X-Ray pulsar output). Therefore long-life SGR
(linked to anomalous X-ray AXPs) may be repeating; if they are
around our galaxy they might be observed again as the few known
ones and the few rare extragalactic XRFs. The orientation of the
beam respect to the line of sight plays a key role in
differentiating the wide GRB morphology. The relativistic cone is
as small as the inverse of the electron progenitor Lorentz factor.
To observe the inner beamed GRB events, one needs the widest SN
sample and the largest cosmic volumes. Therefore the most far away
are usually the brightest. On the contrary, the nearest ones,
within tens Mpc distances, are mostly observable on the cone jet
\textit{periphery}, a bit off-axis. Their consequent large impact
crossing angle leads to longest \textit{anomalous} SN-GRB
duration, with lowest fluency and the softest spectra, as in
earliest GRB98425 and in particular recent GRB060218 signature. A
majority of GRB jet blazing much later (weeks, months after their
SN) may hide their progenitor explosive after-glow and therefore
they are called \textit{orphan} GRB. Conical shape of few nebulae
and the precessing jet of few known micro-quasar, describe in
space the model signature. Recent
outstanding episode of X-ray precursor, ten minutes before the
main GRB event, cannot be understood otherwise.

In our model to make GRB-SN in nearly energy equipartition the jet
must be very collimated $\frac{\Omega}{\Delta\Omega}\simeq
10^{8}$-$10^{10}$ (\cite{FaSa95b, Fa99, DaF05}) explaining why
apparent (but beamed) GRB luminosity $\dot{E}_{GR-jet}\simeq
10^{53}$-$10^{54}$ erg $s^{-1}$ coexist on the same place and
similar epochs with lower (isotropic) SN powers
$\dot{E}_{SN}\simeq 10^{44}-10^{45} erg s^{-1}$. In order to fit
the statistics between GRB-SN rates, the jet must have a decaying
activity ($\dot{L}\simeq (\frac{t}{t_o})^{-\alpha}$, $\alpha
\simeq 1$): it must survive not just for the observed GRB duration
but for a much longer timescale $t_o$, possibly thousands of time longer observed GRB signal
$t_o\simeq10^4\,s$. The much late stages of the GRBs (within the same
decaying power law) would appear as a SGRs: indeed the same law
for GRB output at late time (thousand years) is still valid for
SGRs active at X-ray pulsar Luminosity. SGRs are not Magnetar fire-ball explosion but blazing jets
with lower power (than GRB) but also much nearer distances.
\section{The puzzle of a huge SGR1806-20 flare and the GRB-SGR connection }
The SGR are also puzzling (within explosive Magnetar model).
Why SGR1806-20 of 2004 Dec. 27th, shows no evidence of the loss of its
period $P$ or its derivative $\dot{P}$ after the huge
\textit{Magnetar} eruption, while in this model its hypothetical
magnetic energy reservoir (linearly proportional to
$P\cdot\dot{P}$) must be largely exhausted? Why do SGR1806 radio
afterglows show after its huge eruption 2004 a mysterious two-bump radio curve implying
additional energy injection many days later? Why its wide polarization change
took place in days after the explosion. A precessing thin jet may explain all the puzzles. Why has the SGR1806-20
polarization curve been changing angle radically in short ($\sim$
days) timescale?
Indeed the puzzle (for one shot popular Magnetar-Fireball
model\cite{DuTh92}) arises for the surprising giant flare from SGR
1806-20 that occurred on 2004 December 27th: if it has been
radiated isotropically (as assumed by the Magnetar
model\cite{DuTh92}), most of - if not all - the magnetic energy
stored in the neutron star NS, should have been consumed at once.
This should have been reflected into sudden angular velocity loss
(and-or its derivative) which was \textit{never observed}. On the
contrary a thin collimated precessing jet $\dot{E}_{SGR-jet}\simeq
10^{36}$-$10^{38}$ erg $s^{-1}$, blazing on-axis, may be the
source of such an apparently (the inverse of the solid beam angle
$\frac{\Omega}{\Delta\Omega}\simeq10^{8}$-$10^{9}$) huge bursts
$\dot{E}_{SGR-Flare}\simeq10^{38}\cdot\frac{\Omega}{\Delta\Omega}\simeq10^{47}$
erg $s^{-1}$ with a moderate  steady jet output power (X-Pulsar,
SS433). This explains the absence of any variation in the
SGR1806-20 period and its time derivative, contrary to any obvious
correlation with the dipole energy loss law.

In our model, the temporal evolution of the angle between the
spinning (PSRs), precessing (binary, nutating) jet direction and
the rotational axis of the NS, can be expressed as
\[
\theta_1(t)=\sqrt{\theta_x^2+\theta_y^2}
\]
where
$$\theta_y(t)=
\theta_a\cdot\sin\omega_0t+\cos(\omega_bt+\phi_{b})+\theta_{psr}\cdot\cos(\omega_{psr}t+\phi_{psr})\cdot|(\sin(\omega_Nt+\phi_N))|+
$$
$$
+\theta_s\cdot\cos(\omega_st+\phi_{s})+\theta_N\cdot\cos(\omega_Nt+\phi_N))+\theta_y(0)
$$
and a similar law express the $\theta_x(t)$ evolution. The angular
velocities and phase labels are self-explained\cite{DaF05, DaF06}.
Lorentz factor $\gamma$ of the jet's relativistic particles, for
the most powerful SGR1806-20 event, and other parameters adopted
for the jet model represented in Fig. \ref{fig2} are shown in (\cite{DaF05, DaF06}).

The simplest  way to produce the $\gamma$ emission  would be by IC
of GeVs electron pairs onto thermal infra-red photons. Also
electromagnetic showering of PeV electron pairs by synchrotron
emission in galactic fields, ($e^{\pm}$ from muon decay) may be
the progenitor of the $\gamma$ blazing jet. However, the main
difficulty for a jet of GeV electrons is that their propagation
through the SN radiation field is highly suppressed. UHE muons
($E_{\mu}\geq$ PeV) instead are characterized by a longer
interaction length either with the circum-stellar matter and the
radiation field, thus they have the advantage to avoid the opacity
of the star and escape the dense GRB-SN isotropic radiation field
\cite{DaF05, DaF06}. We propose that also the emission of SGRs is
due to a primary hadronic jet producing ultra relativistic
$e^{\pm}$ (1 - 10 PeV) from hundreds PeV pions,
$\pi\rightarrow\mu\rightarrow e$, (as well as EeV neutron decay in
flight): primary protons can be accelerated by the large magnetic
field of the NS up to EeV energy. The protons could in principle
emit directly soft gamma rays via synchrotron radiation with the
galactic magnetic field
($E_{\gamma}^p\simeq10(E_p/EeV)^2(B/2.5\cdot10^{-6}\,G)$ keV), but
the efficiency is poor because of the too small proton
cross-section, too long timescale of proton synchrotron
interactions. By interacting with the local galactic magnetic
field relativistic pair electrons lose energy via synchrotron
radiation:
$E_{\gamma}^{sync}\simeq4.2\cdot10^6(\frac{E_e}{5\cdot10^{15}\,eV})^2(\frac{B}{2.5\cdot10^{-6}\,G})\,eV$
with a characteristic timescale
$t^{sync}\simeq1.3\cdot10^{10}(\frac{E_{e}}{5\cdot10^{15}\,eV})^{-1}(\frac{B}{2.5\cdot10^{-6}\,G})^{-2}\,s$.
This mechanism would produce a few hundreds keV radiation as it is
observed in the intense $\gamma$-ray flare from SGR 1806-20.

The Larmor radius is about two orders of magnitude smaller than
the synchrotron interaction length and this may imply that the
aperture of the showering jet is spread in a fan structure
\cite{Fa97, Fa00-04} by the magnetic field,
$\frac{R_L}{c}\simeq4.1\cdot10^{8}(\frac{E_{e}}{5\cdot10^{15}\,eV})(\frac{B}{2.5\cdot10^{-6}\,G})^{-1}\,s$.
Therefore the solid angle is here the inverse of the Lorentz
factor ($\sim$ nsr). In particular a thin
($\Delta\Omega\simeq10^{-9}$-$10^{-10}$ sr) precessing jet from a
pulsar may naturally explain the negligible variation of the spin
frequency $\nu=1/P$ after the giant flare ($\Delta\nu<10^{-5}$
Hz). Indeed it seems quite unlucky that a huge
($E_{Flare}\simeq5\cdot10^{46}$ erg) explosive event, as the
needed mini-fireball by a magnetar model\cite{DuTh92}, is not
leaving any trace in the rotational energy of the SGR 1806-20, $
E_{rot}=\frac{1}{2}I_{NS}\omega^2\simeq3.6\cdot10^{44}(\frac{P}{7.5\,s})^{-2}(\frac{I_{NS}}{10^{45}g\,cm^2})$
erg. The consequent fraction of energy lost after the flare is
severely bounded by observations:
$\frac{\Delta(E_{Rot})}{E_{Flare}}\leq10^{-6}$. More absurd in
Magnetar-explosive model is the evidence of a brief precursor
event (one-second SN output) taking place with no disturbance on
SGR1806-20 \textit{two minutes before} the hugest flare of 2004
Dec. 27th. The thin precessing Jet while being extremely
collimated (solid angle
$\frac{\Omega}{\Delta\Omega}\simeq10^{8}$-$10^{10}$
(\cite{FaSa95b, Fa99, DaF05, DaF06}) may blaze at different angles
within a wide energy range (inverse of
$\frac{\Omega}{\Delta\Omega}\simeq10^{8}$-$10^{10}$). The output
power may exceed $\simeq10^{8}$, explaining the extreme low
observed output in GRB980425 -an off-axis event-, the long late
off-axis gamma tail by  GRB060218\cite{Fargion-GNC}),  respect to
the on-axis and more distant GRB990123 (as well as GRB050904).
\section{ Conclusion: Hundred GeVs $\gamma$, $\nu_{\mu}$ and PeVs  $\nu_{\tau}$ precursors}
 The GRBs are not the most powerful
explosions, but just the most  collimated ones. Their birth rate
is comparable to the SN ones (a few a second in the observable
Universe), but their thin beaming ($10^{-8}$ sr) make them extremely
rare ($10^{-8} s^{-1}$ ) rate to point to us at their very birth.
 The persistent precessing (slow decay of scale time of hours)
 and moving beam span a wider angle with time and
 it encompass a larger solid angle increasing the rate by 3 order of
 magnitude to observed GRB rate; after a few hours $\simeq 10^4 s.$
the beam may hit the Earth and appear as a GRB near coincident
with a SN.  The power law decay mode of the jet make it alive at
smaller power days, months and year later, observable only at
nearer and middle distance as a Short GRB or (at its jet
periphery) as an XRF or in our galaxy as a SGRs. The presence of a huge population of active jets fit a
wide spectrum of GRB morphology \cite{Giovannelli}.  Now in our Universe
thousands of GRBs are shining at SN peak power, but pointing else
where. Only one a day might be blazing to us and captured at SWIFT
threshold level. Thousand  of billions  are blazing (unobserved)
as SGRs in the Universe. Short GRBs as well SGRs are born in SNRs
location and might be revealed in nearby spaces. The GRB-SGRs
connection with XRay-Pulsars make a possible link to AXRay pulsar
jets recently observed in most X-gamma sources as the famous Crab.
The possible GRB-SGR link to X-gamma pulsar is a natural
possibility to be considered as a grand unification of the model.
Our prediction is that a lower threshold, as GLAST-Fermi satellite will
induce a higher rate of GRBs both at nearer volumes (as GRB060218
and GRB 980425) and at largest red-shifts. Rarest hard GeV $\gamma$ at TeV are too far and opaque by IR cut-off. Rarer Tens GeV neutrino GRB muons may trace in $Km^3$ detector the direction of such far GRB whose prompt X-ray afterglow maybe followed up. In analogy up-ward tens PeVs Tau air-shower may trigger the prompt X-Ray telescope searching afterglows of far hidden GRBs. Their discovery will confirm the PeV-SN-GRB-Jet model.\\

\end{document}